\documentclass[apjl]{emulateapj}
\usepackage{graphicx}
\usepackage{ifpdf}
\usepackage{ifthen}

\newboolean{emulateapj}
\setboolean{emulateapj}{true}

\newcommand{\ctbd}[1]{}

\newcommand{\lc}{light curve}
\newcommand{\lcs}{light curves}
\newcommand{\Lc}{Light curve}

\newcommand{\cfa}{Harvard-Smithsonian Center for Astrophysics (CfA)}

\newcommand{\kms}{\ensuremath{\rm km\,s^{-1}}}
\newcommand{\ms}{\ensuremath{\rm m\,s^{-1}}}
\newcommand{\mss}{\ensuremath{\rm m\,s^{-2}}}
\newcommand{\gcmc}{\ensuremath{\rm g\,cm^{-3}}}

\newcommand{\hd}[1]{\mbox{HD #1}}

\newcommand{\teff}{\ensuremath{T_{\rm eff}}}
\newcommand{\teq}{\ensuremath{T_{\rm eq}}}
\newcommand{\logg}{\ensuremath{\log{g}}}
\newcommand{\vsini}{\ensuremath{v \sin{i}}}
\newcommand{\feh}{[Fe/H]}

\newcommand{\rsun}{\ensuremath{R_\sun}}
\newcommand{\msun}{\ensuremath{M_\sun}}
\newcommand{\lsun}{\ensuremath{L_\sun}}

\newcommand{\rstar}{\ensuremath{R_\star}}
\newcommand{\mstar}{\ensuremath{M_\star}}
\newcommand{\lstar}{\ensuremath{L_\star}}

\newcommand{\rpl}{\ensuremath{R_{p}}}
\newcommand{\mpl}{\ensuremath{M_{p}}}

\newcommand{\rhopl}{\ensuremath{\rho_{p}}}

\newcommand{\gpl}{\ensuremath{g_{p}}}
\newcommand{\fpl}{\ensuremath{f_{p}}}

\newcommand{\rjup}{\ensuremath{R_{\rm J}}}
\newcommand{\mjup}{\ensuremath{M_{\rm J}}}

\newcommand{\rjuplong}{\ensuremath{R_{\rm Jup}}}
\newcommand{\mjuplong}{\ensuremath{M_{\rm Jup}}}

\newcommand{\figr}[1]{Fig.~\ref{fig:#1}}
\newcommand{\secr}[1]{\mbox{\S\ \ref{sec:#1}}}

\newcommand{\tabr}[1]{\mbox{Table~\ref{tab:#1}}}

\newcommand{\flwof}{\mbox{FLWO 1.2 m}}

\newcommand{\wom}{\mbox{Wise 1 m}}

\newcommand{\woc}{\mbox{Wise 0.46 m}}

\newcommand{\hatcur}{HAT-P-9}
\newcommand{\hatcurb}{HAT-P-9b}

\newcommand{\datapts}{\ensuremath{6884}} 
\newcommand{\hatcurm}{\ensuremath{0.78\pm0.09}}
\newcommand{\hatcurr}{\ensuremath{1.40\pm0.06}}
\newcommand{\hatcurrho}{\ensuremath{0.35\pm0.06}}
\newcommand{\hatcurimp}{\ensuremath{0.52\pm0.03}}
\newcommand{\hatcurg}{\ensuremath{9.8\pm1.0}} 
\newcommand{\hatcurTeq}{\ensuremath{1530\pm40}}
\newcommand{\hatcurSaf}{\ensuremath{0.046\pm0.007}}
\newcommand{\hatcurfp}{\ensuremath{1.3^{+0.3}_{-0.3}}}
\newcommand{\hatcurar}{\ensuremath{8.6\pm0.2}}
\newcommand{\hatcurrprstar}{\ensuremath{0.1083\pm0.0005}}
\newcommand{\hatcurarel}{\ensuremath{0.053\pm0.002}}
\newcommand{\hatcurP}{\ensuremath{3.92289\pm0.00004}}
\newcommand{\hatcurK}{\ensuremath{84.7\pm7.9}}
\newcommand{\hatcuri}{\ensuremath{86.5\pm0.2}}
\newcommand{\hatcurgamma}{\ensuremath{22.665\pm0.006}}
\newcommand{\hatcurT}{\ensuremath{2,\!454,\!417.9077\pm0.0003}}
\newcommand{\hatcurdur}{\ensuremath{0.143\pm0.004}}
\newcommand{\hatcuringdur}{\ensuremath{0.019\pm0.003}}

\newcommand{\hatcurra}{\ensuremath{07^{\mathrm{h}}\, 20^{\mathrm{m}}\, 40.^{\mathrm{\! \! s}}44}}
\newcommand{\hatcurdec}{\ensuremath{+37\arcdeg\, 08\arcmin\ \  26.{\! \arcmin}3}}
\newcommand{\hatcurms}{\ensuremath{1.28\pm0.13}}
\newcommand{\hatcurrs}{\ensuremath{1.32\pm0.07}}
\newcommand{\hatcurloggs}{\ensuremath{4.29^{+0.03}_{-0.04}}}
\newcommand{\hatcurages}{\ensuremath{1.6^{+1.8}_{-1.4}}}
\newcommand{\hatcurMV}{\ensuremath{3.7^{+0.3}_{-0.2}}}
\newcommand{\hatcurBV}{\ensuremath{0.50^{+0.06}_{-0.05}}}

\newcommand{\hatcurteff}{\ensuremath{6350\pm150}}
\newcommand{\hatcurfeh}{\ensuremath{0.12\pm0.20}}
\newcommand{\hatcurdist}{\ensuremath{480\pm60}}
\newcommand{\hatcurvsini}{\ensuremath{11.9\pm1.0}}	
\newcommand{\hatcurmV}{\ensuremath{12.297\pm 0.063}}
\newcommand{\hatcurlogls}{\ensuremath{0.41_{-0.09}^{+0.08}}}

\shorttitle{\hatcurb}
\shortauthors{Shporer et al.}

\begin{document}
\ifthenelse{\boolean{emulateapj}}{
\title{\hatcurb: A Low Density Planet Transiting a Moderately Faint F star \altaffilmark{1}}}
{\title{\hatcurb: A Low Density Planet Transiting a Moderately Faint F star \altaffilmark{1}}}
\author{
	Avi~Shporer\altaffilmark{2},
	G\'asp\'ar~\'A.~Bakos\altaffilmark{3,4},
	Francois~Bouchy\altaffilmark{5},
	Frederic~Pont\altaffilmark{6},
	G\'eza~Kov\'acs\altaffilmark{7},
	Dave~W.~Latham\altaffilmark{3},
	Brigitta~Sip\H{o}cz\altaffilmark{8,3},
	Guillermo~Torres\altaffilmark{3},
	Tsevi~Mazeh\altaffilmark{2},
	Gilbert~A.~Esquerdo\altaffilmark{3,9},
	Andr\'as~P\'al\altaffilmark{3,8},
	Robert~W.~Noyes\altaffilmark{3},
	Dimitar~D.~Sasselov\altaffilmark{3},
	J\'ozsef~L\'az\'ar\altaffilmark{10},
	Istv\'an~Papp\altaffilmark{10},
	P\'al~S\'ari\altaffilmark{10} \&
	G\'abor~Kov\'acs\altaffilmark{3}
}
\altaffiltext{1}{
	Based in part on radial velocities obtained with the SOPHIE
	spectrograph mounted on the 1.93m telescope at
	the Observatory of Haute Provance (runs 07A.PNP.MAZE,
	07B.PNP.MAZE, 08A.PNP.MAZE).
}
\altaffiltext{2}{Wise Observatory, Tel Aviv University, Tel Aviv,
    Israel 69978; shporer@wise.tau.ac.il.}
\altaffiltext{3}{\cfa, 60 Garden Street, Cambridge, MA 02138, USA.}
\altaffiltext{4}{NSF Fellow.}
\altaffiltext{5}{Institut d'Astrophysique de Paris, 98bis Bd Arago,
	75014 Paris, France.}
\altaffiltext{6}{School of Physics, University of Exeter, Stocker Road, Exeter EX4 4QL, United Kingdom.}
\altaffiltext{7}{Konkoly Observatory, Budapest, P.O.~Box 67, H-1125, Hungary.}
\altaffiltext{8}{Department of Astronomy,
	E\"otv\"os Lor\'and University, Pf.~32, H-1518 Budapest, Hungary.}
\altaffiltext{9}{Planetary Science Institute, 620 N. 6th Avenue, Tucson, Arizona
 85705.}
\altaffiltext{10}{Hungarian Astronomical Association, 1461 Budapest,
	P.~O.~Box 219, Hungary.}

\setcounter{footnote}{10}

\begin{abstract}
	We report the discovery of a planet transiting a moderately faint
	($V=12.3$\,mag) late F star, with an orbital period of \hatcurP\,days.
	From the transit light curve and radial velocity measurements
	we determine that the radius of the planet is $\rpl = \hatcurr\,\rjuplong$
	and that the mass is $\mpl = \hatcurm\,\mjuplong$. The density
	of the new planet, $\rhopl = \hatcurrho\,\gcmc$, fits to the
	low-density tail of the currently known transiting planets.
	We find that the center of
	transit is at $T_{\mathrm{c}} = \hatcurT$ (HJD), and the total
	transit duration is \hatcurdur\,days. The host star has $\mstar=\hatcurms\,\msun$
    and $\rstar=\hatcurrs\,\rsun$.
\end{abstract}

\keywords{
	stars: individual: {\mbox GSC 02463-00281} \---
	planetary systems: individual: \hatcurb
}

\section{Introduction}
\label{sec:intro}

Transiting extra-solar planets are important astrophysical objects as
they allow to test planetary structure and evolution theory
\citep[e.g.,][]{fortney08a, burrows08, baraffe08} especially because they yield a
measurement of the planetary mass and radius. Over the last few years, the
sample of known transiting planets has grown substantially, leading
to an improved theoretical understanding
of their physical nature \citep[e.g.,][]{Guillot06, burrows07, fortney08b, chabrier07}.
In addition, the increasing
sample of transiting planets enabled the discovery
of some interesting correlations, such as the mass-period
relation \citep{mazeh05, gaudi05}. The astrophysics behind these correlations
is not fully understood, and a clear way towards progress is the
discovery of many more transiting planets.

We report here the discovery of another transiting planet detected by the
HATNet project\footnote{http://www.hatnet.hu} \citep{bakos02,bakos04},
labeled \hatcurb, and our determination of its
parameters, including mass, radius, density and surface gravity.
In \secr{anal} we describe our photometric and spectroscopic observations
and in \secr{stelpar} we derive the stellar parameters. The orbital solution
is performed in \secr{rv} and a discussion of possible blend scenarios is brought
in \secr{blend}. The determination of the \lc\ parameters and the planet's
physical parameters is described in \secr{planpar} and we bring a discussion
in \secr{disc}.

\section{Observations and Analysis}
\label{sec:anal}

\subsection{Detection of the transit in the HATNet data}
\label{sec:dete}

\ifthenelse{\boolean{emulateapj}}{\begin{figure}[t]}{\begin{figure}[t]}
\ifpdf
\plotone{f1.pdf}
\else
\plotone{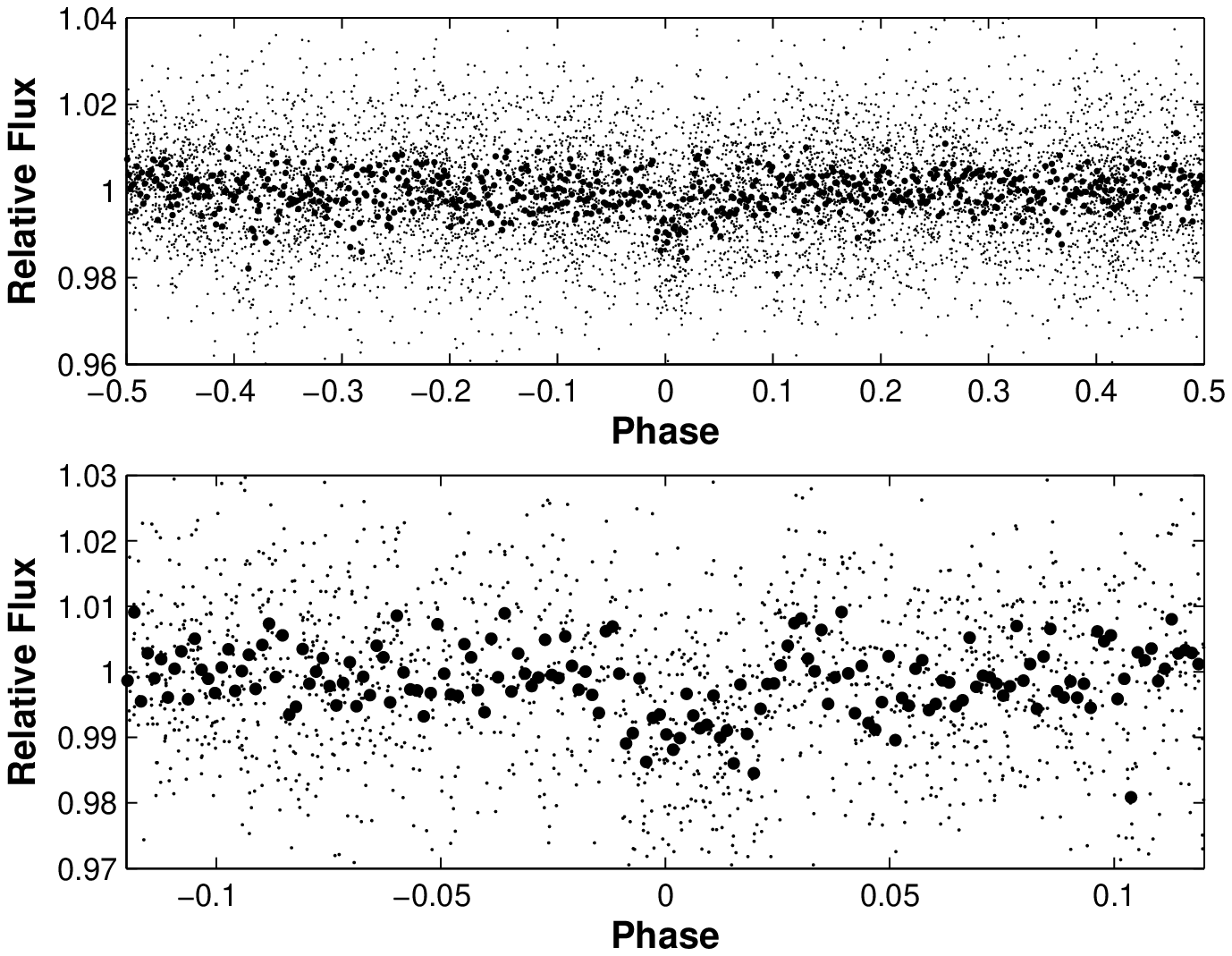}
\fi
\caption{
	The top panel shows the unbinned HATNet \lc\ with \datapts\ data points,
	phased with the orbital period of $P=3.92289$\,d. The binned \lc\
	is over-plotted. The center of the transit is at phase zero.
	The bottom panel shows a zoom-in around the transit.
	The HATNet \lc\ is folded with the ephemeris obtained here, based on
	the follow-up \lcs\ (see \secr{photfol}). The small offset between
	phase zero and transit center suggests there is a slight difference in
	the ephemeris for the HATNet data and the photometric follow-up
	data, obtained $\sim 1000$ days apart. See \secr{disc} for further
	discussion.
\label{fig:lc1}}
\ifthenelse{\boolean{emulateapj}}{\end{figure}}{\end{figure}}

\hatcur\ is positioned in HATNet's internally labeled field G176,
centered at
$\alpha=07^{\mathrm{h}}28^{\mathrm{h}}$, $\delta = 37\arcdeg 30\arcmin$.
This field was observed in network mode by the HAT-6 telescope,
located at the Fred Lawrence Whipple Observatory (FLWO) of the Smithsonian
Astrophysical Observatory
(SAO), and the HAT-9 telescope at the Submillimeter Array (SMA) site atop
Mauna Kea, Hawaii. In total, \datapts\ exposures were obtained at a 5.5
minute cadence between 2004 November 26 and 2005 October 21 (UT).

Preliminary reduction included standard bias, dark and flat-field
corrections, followed by an astrometric solution using the code
of \citet{pal06}. Photometry was applied using fine-tuned aperture
photometry while the raw \lcs\ were processed by our external
parameter decorrelation technique. Next, the Trend Filtering
Algorithm \citep[TFA;][]{kovacs05} was applied to get the final \lcs. To search for the
signature of a transiting planet in the \lcs\ we used the Box Least
Squares \citep[BLS;][]{kovacs02} algorithm.

A transit-like signal was
identified in the \lc\ of \hatcur\ (GSC 02463-00281, 2MASS
J07204044+3708263), which is a $V$=12.3\,mag star,
fainter than most transiting planet host stars detected with small-aperture,
wide-field, ground-based campaigns. The HATNet \lc\ is presented at the top
panel of \figr{lc1}, folded on the ephemeris obtained from analyzing
the follow-up \lcs\ ($P$ = 3.92289\,days and $T_{\rm c} = 2454417.9077$,
see \secr{photfol}), showing a flux
decrement of $\sim1\,\%$ at phase zero. The figure shows a small shift
between phase zero and transit center, suggesting a possible slight difference
in the ephemeris for the HATNet data and the photometric follow-up data,
obtained $\sim 1000$ days apart (see \secr{disc} for further discussion).
This transiting planet candidate, along with others from the same field,
was selected for follow-up observations, to investigate its nature.

\subsection{Early spectroscopy follow-up}
\label{sec:ds}

Initial follow-up observations were made with the CfA Digital
Speedometer \citep[DS;][]{latham92} in order to characterize the host
star and to reject obvious astrophysical false-positive scenarios that
mimic planetary transits. The five radial velocity (RV) measurements
obtained over an interval of 62 days showed an rms residual of
1.1\,\kms, consistent with no detectable RV variation. Atmospheric
parameters for the star (effective temperature $T_{\rm eff}$, surface
gravity $\log g$, and projected rotational velocity
$v \sin i$) were derived as described by \cite{torres02}, initially
assuming a fixed metallicity of [Fe/H] $= 0.0 \pm 0.2$. We obtained
$\log g = 4.05 \pm 0.30$,
$\teff = 6130 \pm 150$\,K and $\vsini = 12.2 \pm 1.0$\,\kms.
As the DS results were consistent with a planet orbiting a
moderately-rotating main sequence star, this target was selected
for high-precision spectroscopy follow-up.

\subsection{High-precision spectroscopy follow-up}
\label{sec:hires}

\ifthenelse{\boolean{emulateapj}}{\begin{figure}[t]}{\begin{figure}[t]}
\ifpdf
\plotone{f2.pdf}
\else
\plotone{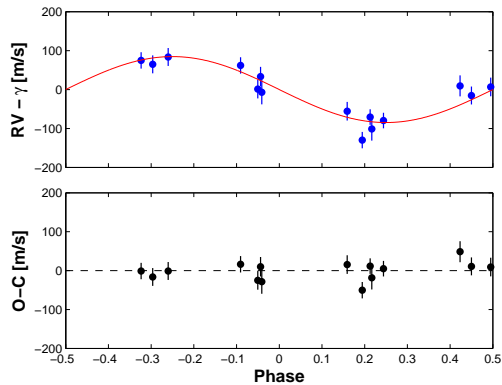}
\fi
\caption{
	The top panel shows the RV measurements phased with the period of
	$P=3.92289$\,days and mid-transit time of $T_{\rm c} = 2454417.9077$ (HJD). 
    The zero-point in phase corresponds to the epoch
	of mid-transit. Overlaid is the best sinusoidal fit.
	The bottom panel shows the residuals from the fit.
\label{fig:rv1}}
\ifthenelse{\boolean{emulateapj}}{\end{figure}}{\end{figure}}

\begin{deluxetable}{cccrc}
\tabletypesize{\scriptsize}
\tablewidth{0pt}
\tablecaption{
	\label{tab:rv}
	Radial Velocities for \hatcur.
}
\tablewidth{0pt}
\tablehead{
	\colhead{BJD $- 2,\!400,\!000$} &
	\colhead{RV\tablenotemark{a}} &
	\colhead{$\sigma_{\rm RV}$} &
	\colhead{BS\tablenotemark{b}} &
	\colhead{S/N\tablenotemark{c}} \\
	\colhead{(days)} &
	\colhead{(\ms)} &
	\colhead{(\ms)} &
	\colhead{(\ms)} &
}
\startdata
54233.3347 &   22667 &24   &  -62$\ \ \ \ $& 43 \\
54379.6357 &   22586 &20   &    5$\ \ \ \ $& 44 \\
54380.6213 &   22672 &24   &  -93$\ \ \ \ $& 38 \\
54381.5815 &   22749 &23   &  -148$\ \ \ \ $&40 \\
54413.6292 &   22727 &21   &    3$\ \ \ \ $& 43 \\
54414.6084 &   22610 &24   &  -31$\ \ \ \ $& 38 \\
54415.6457 &   22675 &27   &  -45$\ \ \ \ $& 34 \\
54416.6392 &   22740 &21   &  -13$\ \ \ \ $& 42 \\
54421.6701 &   22659 &31   &  -11$\ \ \ \ $& 32 \\
54422.6651 &   22595 &20   &  -59$\ \ \ \ $& 46 \\
54587.3543 &   22536 &21   &   10$\ \ \ \ $& 44 \\
54588.3571 &   22650 &23   &   30$\ \ \ \ $& 41 \\
54589.3525 &   22730 &23   &   -2$\ \ \ \ $& 40 \\
54590.3425 &   22699 &25   &  -45$\ \ \ \ $& 38 \\
54591.3647 &   22564 &30   &   10$\ \ \ \ $& 32

\enddata
\tablenotetext{a}{The RVs include the barycentric correction.}
\tablenotetext{b}{Bisector span.}
\tablenotetext{c}{Signal to noise ratio per pixel at $\lambda = 5500$\,\AA.}
\label{SOPHIE}
\end{deluxetable}

Observations were carried out at the Haute
Provence Observatory (OHP) 1.93-m telescope, with the SOPHIE
spectrograph \citep{bouchy06}. SOPHIE is a multi-order echelle
spectrograph fed through two fibers, one of which is used for starlight
and the other for sky background or a wavelength calibration lamp. The
instrument is entirely computer-controlled and a standard data
reduction pipeline automatically processes the data upon CCD readout.
RVs are calculated by numerical cross-correlation with a high
resolution observed spectral template of a G2 star.

\hatcur\ was observed with SOPHIE in the high-efficiency mode (R $\sim 39000$)
during three observing runs, from 2007 May until 2008 May.
Due to the relative faintness of this star,
exposure times were in the range of 25 to 75 minutes, depending on
observing conditions. The resulting
signal to noise ratios were 32--46 per pixel at $\lambda = 5500$\,\AA.
Using the empirical relation of \citet{cameron07} we estimated the
RV photon-noise uncertainties to be 20--31\,\ms. We made 17 RV measurements
in total. Two of those were highly contaminated by the Moon and were
ignored, leaving 15 RV measurements, listed in Table~\ref{SOPHIE}.

\subsection{Photometry follow-up}
\label{sec:photfol}

\ifthenelse{\boolean{emulateapj}}{\begin{figure}[t]}{\begin{figure}[t]}
\ifpdf
\plotone{f3.pdf}
\else
\plotone{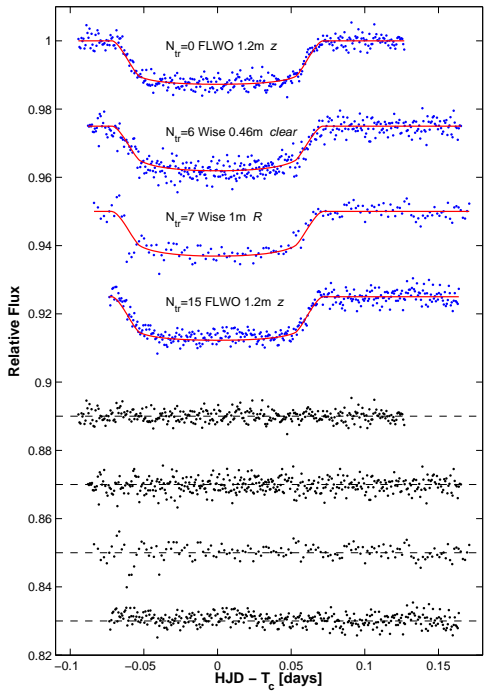}
\fi
\caption{
    Follow-up \lcs\ of \hatcur. The four \lcs\ are shown at the top
	of the figure, with the fitted model over-plotted. The label next
	to each \lc\ gives the transit event number, observatory, telescope
	and filter used. Residuals are presented at the bottom of the
	figure, with the same top to bottom order as the actual \lcs.
\label{fig:lc2}}
\ifthenelse{\boolean{emulateapj}}{\end{figure}}{\end{figure}}

\notetoeditor{This is the intended place of \tabr{lcsum}}
\begin{deluxetable*}{lllcccccc}
\tabletypesize{\scriptsize}
\tablecaption{
	List of follow-up \lcs\ of \hatcur.
\label{tab:lcsum}}
\tablehead{
        \colhead{$N_{\mathrm{tr}}$}     &
	\colhead{Start Date}  &
	\colhead{Observatory+}&
        \colhead{Filter}     &
	\colhead{$u_1$}      &
	\colhead{$u_2$}      &
	\colhead{Cadence} &
	\colhead{$\beta$\tablenotemark{a}} &
	\colhead{RMS}  \\
	\colhead{}&
	\colhead{UT}&
	\colhead{Telescope}&
	\colhead{}&
	\colhead{}&
	\colhead{}&
	\colhead{min$^{-1}$}&
	\colhead{}&
    \colhead{$\%$}
}
\startdata
\hspace{4pt}0  & 2007 Nov 13             & FLWO 1.2 m  & $z$     & 0.1313 & 0.3664 & 1.0 & 1.1 & 0.17\\
\hspace{4pt}6  & 2007 Dec \hspace{5pt}6  & Wise 0.46 m & clear   & 0.2403 & 0.3816 & 1.1 & 1.0 & 0.21\\
\hspace{4pt}7  & 2007 Dec \hspace{1pt}10 & Wise 1.0 m  & $R$     & 0.2403 & 0.3816 & 0.4 & 1.2 & 0.22\\
15             & 2008 Jan \hspace{2pt}11 & FLWO 1.2 m  & $z$     & 0.1313 & 0.3664 & 1.0 & 1.6 & 0.18

\enddata
\tablenotetext{a}{The correlated noise factor, by which the errors
of each \lc\ are multiplied (see \secr{photfol}).
}
\end{deluxetable*}

In order to better characterize the transit parameters and to
derive a better ephemeris, we performed photometric follow-up observations
with 1-m class telescopes. We obtained a total of four transit \lcs\
of \hatcurb, shown in \figr{lc2}. Two events were observed by the
KeplerCam detector on the \flwof\ telescope \citep[see][]{holman07} on
UT 2007 Nov 13 and UT 2008 Jan 11, in the Sloan $z$-band. We refer to the
2007 Nov 13 event as having a transit number $N_{\mathrm{tr}}=0$, so the
2008 Jan 11 transit number is $N_{\mathrm{tr}}=15$. In addition, two \lcs\
were obtained at the Wise Observatory. On UT 2007 Dec 6 the
$N_{\mathrm{tr}}=6$ transit event was observed by the \woc\ telescope
\citep{brosch08}, with no filter. The following event, with
$N_{\mathrm{tr}}=7$, was observed on UT 2007 Dec 10 by the \wom\ telescope
in the Cousins $R$-band. An additional \lc, obtained with
the \flwof\ telescope, was of poor quality and is not
included here. \tabr{lcsum} lists for each \lc\ its $N_{\mathrm{tr}}$
number, UT date, observatory, telescope, filter used, limb-darkening
coefficients used in its analysis, mean cadence, the correlated noise
$\beta$ factor (see below) and the RMS residuals from the fitted \lc\ model.

Data were reduced in a similar manner to the HATNet data,
using aperture photometry and an ensemble of $\sim$100 comparison stars
in the field. An analytic model was fitted to these data, as described below
in \secr{planpar}, and yielded a period of \hatcurP\,d and a reference
epoch of mid-transit $T_{\mathrm{c}}=\hatcurT$\,d (HJD). The length of
the transit as determined from this joint fit is \hatcurdur\,d (3 hours,
26 minutes), the length of ingress is \hatcuringdur\,d (27 minutes),
and the central transit depth is $1.17 \pm 0.01$\,\%. The latter value
is simply the square of the radius ratio (see \tabr{orb}) if one ignores
the limb-darkening effect. This effect increases the depth by
about $0.1$\,\%, in the bands used here.

\section{Stellar parameters}
\label{sec:stelpar}

\notetoeditor{This is the intended place of \tabr{stelpar}}
\begin{deluxetable}{lll}
\tabletypesize{\scriptsize}
\tablecaption{
	Summary of stellar parameters for \hatcur.
\label{tab:stelpar}}
\tablehead{
	\colhead{Parameter} &
	\colhead{Value} &
	\colhead{Source}
}
\startdata
R.A.          &  \hatcurra                  & 2MASS\\
Dec.          &  \hspace{-6.5pt}\hatcurdec  & 2MASS\\
$m_V$ (mag)   &  \hatcurmV                  & TASS \\
\teff\ (K)    &  \hatcurteff	            & DS\,+\,Yonsei-Yale \\
\vsini\ (\kms)&  \hatcurvsini               & DS\,+\,Yonsei-Yale \\
\logg 	      &  \hatcurloggs               & DS\,+\,Yonsei-Yale\,+\,light curve shape\\
\feh\ (dex)   &  \hatcurfeh                 & DS\,+\,Yonsei-Yale \\
Mass (\msun)  &  \hatcurms                  & Yonsei-Yale\,+\,light curve shape  \\
Radius (\rsun)&  \hatcurrs                  & Yonsei-Yale\,+\,light curve shape  \\
$\log (\lstar /\lsun)$&\hatcurlogls         & Yonsei-Yale \\
$M_V$ (mag)   &  \hatcurMV	                & Yonsei-Yale \\
Age (Gyr)     &  \hatcurages                & Yonsei-Yale\,+\,light curve shape  \\
B-V (mag)     &  \hatcurBV                  & Yonsei-Yale \\
Distance (pc)\tablenotemark{a}&\hatcurdist  & Yonsei-Yale
\enddata
\tablenotetext{a}{Assuming extinction of $A(V) = 0.15$\,mag,
see text at \secr{stelpar}.}
\end{deluxetable}

The mass (\mpl) and radius (\rpl) of a transiting planet, determined 
from transit photometry and RV data, is dependent also on those of 
the parent star. In order to determine the stellar properties
needed to place \mpl\ and \rpl\ on an absolute scale, we made use of
stellar evolution models along with the observational constraints from
spectroscopy and photometry, as described in \citet{torres08}.
Because of its relative faintness, the host star does not
have a parallax measurement from {\it Hipparcos\/}, and thus a direct
estimate of the absolute magnitude is not available for use as a
constraint. An alternative approach is to use the surface gravity of
the star, which is a measure of the evolutionary state of the
star and therefore has a very strong influence on the radius. However,
$\logg$ is a difficult quantity to measure
spectroscopically and is often strongly correlated with other
spectroscopic parameters (see \secr{ds}). It has been pointed out by
\citet{sozzetti07} that the normalized separation of the planet,
$a/\rstar$, can provide a much better constraint for stellar parameter
determination than the spectroscopic \logg.
The $a/\rstar$ quantity can be determined
directly from the photometric observations with no additional
assumptions (other than the values of the parameters describing limb-darkening,
which is a second-order effect), and it is related to the mean density
of the host star. As
discussed later in \secr{planpar}, an analytic fit to the light curve
yields $a/\rstar$ = \hatcurar. This value, along with \teff\
from \secr{ds}, and assuming \feh\ = $0.0 \pm 0.2$,
was compared with the Yonsei-Yale stellar evolution models of
\cite{yi01}.
This resulted in values for the stellar mass and radius of
\mstar = $1.16_{-0.15}^{+0.12}\,\msun$ and \rstar = $1.28 \pm 0.08\,\rsun$,
and an estimated age of $3.6_{-2.2}^{+3.3}$\,Gyr. The result
for \logg\ was \hatcurloggs, consistent with the value
derived from the DS spectra.

Nevertheless, in order to verify the overall consistency of our 
results and refine the stellar parameters, we carried out a new 
iteration. We imposed this latter value of \logg\ (coming from stellar
evolution modeling), and analyzed the DS spectra by
allowing the metallicity, \vsini\ and \teff\ to vary. The new results
were \feh\,=\,\hatcurfeh, \vsini\,=\,\hatcurvsini\,\kms\ and
\teff\,=\,\hatcurteff\,K. Repeating the stellar evolution modeling
resulted in \mstar = \hatcurms\,\msun, \rstar = \hatcurrs\,\rsun\
and an age of \hatcurages\,Gyr.
The stellar properties are summarized in
\tabr{stelpar} and they correspond to a late F star.

We used also our SOPHIE high-resolution spectra to estimate \vsini\
and \feh. Based on our result for $B-V$ from the
stellar evolution models, of $B-V~=~$\hatcurBV\,mag, we got
\vsini\ = $10.1 \pm 1.0$\,\kms\ and \feh = $+0.18 \pm 0.10$\,dex. Those
values are close to the results from the DS spectra and the stellar
evolution model. However, the SOPHIE spectra were taken in
High-Efficiency mode, where there is a known problem
with removing the echelle blaze function. Hence, these estimates
are used only for comparison, and are not included in our final result.

To check our spectroscopically determined \teff\ we used several
publicly available color indices for this star, and the calibrations of
\citet{ramirez05} and \citet{Casagrande06} to derive independent
temperature estimates, using the \feh\ value found above in these
calibrations. We adjusted the reddening till the photometric
temperature matched the spectroscopic temperature, yielding $E(B-V) =
0.053$ \citep{ramirez05} and $E(B-V) = 0.043$ \citep{Casagrande06}. We
adopted the mean value of $0.048$ as reddening, implying an extinction of
$A(V) = 0.15$\,mag. Note that the \citet{burstein82} reddening
maps for this celestial position ($l,b$ = 181.1, 21.4)
give $E(B-V) = 0.072$ and the
\citet{schlegel98} maps yield $E(B-V) = 0.065$,
broadly consistent with our findings, especially if we take into
account that these latter methods measure the total reddening along the
line of sight.

Using the $V$ magnitude from the TASS survey \citep{droege06}, $V$ =
\hatcurmV\,mag, the absolute $V$ magnitude from the stellar
evolution model, and the $A(V) = 0.15$\,mag value determined above,
the distance to \hatcur\ is \hatcurdist\,pc.

\section{Spectroscopic orbital solution}
\label{sec:rv}

Our 15 RV measurements from SOPHIE were fitted with a Keplerian orbit
model solving for the velocity semi-amplitude $K$ and the
center-of-mass velocity $\gamma$, holding the period and transit epoch
fixed at the well-determined values from photometry (see \tabr{orb}).
The eccentricity was initially set to zero.
The fit yields $K=\hatcurK\,\ms$ and $\gamma = \hatcurgamma\,\kms$. The observations
and fitted RV curve are displayed in the top panel of \figr{rv1}.
The residuals are presented in the bottom panel of the same figure.
RMS residuals is 22.1\,\ms, consistent with the RV uncertainties. The
value of $\chi^2$ is 12.4 for 13 degrees of freedom.

\section{Excluding blend scenarios}
\label{sec:blend}

\ifthenelse{\boolean{emulateapj}}{\begin{figure}[t]}{\begin{figure}[t]}
\ifpdf
\plotone{f4.pdf}
\else
\plotone{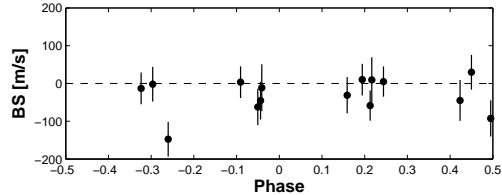}
\fi
\caption{
	Line bisector spans folded on the
	orbital phase. Albeit one or two outliers, the bisector spans do
	not show a variation with the orbital phase.
\label{fig:rv2}}
\ifthenelse{\boolean{emulateapj}}{\end{figure}}{\end{figure}}

We tested the reality of the velocity variations by
examining the spectral line bisector spans (BSs) of the star using our SOPHIE data.
If the measured velocity changes are due only to distortions in the
line profiles arising from contamination of the spectrum by the
presence of a binary with a period of 3.92\,days, we would expect the
BSs (which measure line asymmetry) to vary with this period,
resulting in a correlation
between BS and RV \citep[see, e.g.,][]{queloz01,torres05}. As shown in
\figr{rv2}, the BSs show no significant variations, except one or two
outliers.
The correlation coefficient between BS and
RV for all 15 measurements is -0.38. Ignoring the extreme point reduces
the correlation to -0.18.

In order to estimate the statistical significance of a correlation between
BS and RV we define the following statistics:
\begin{equation}
r_{\sigma} = \frac{\sigma_{BS, fit}}{\sigma_{BS}},
\end{equation}
where $\sigma_{BS}$ is the standard deviation of the BS values
and $\sigma_{BS, fit}$ the standard deviation of the residuals
of a BS fit to the RVs. Both standard deviations are the {\it unbiased}
estimators. For pure noise $r_{\sigma}$ will equal 1.0.
When fitting the BSs to the RVs of all 15 measurements,
using polynomials of degrees 1--3 we got $r_{\sigma}$ in the range 0.984--1.043,
consistent with no significant correlation.

Another sign of a binary would be a dependence between the RV
amplitude and the template used. This may happen when the components
of the blended binary are of a different spectral type than the primary
target. We re-calculated the RVs using F0 and K5 templates and got
an amplitude consistent with our original value, within $1\sigma$ (not
shown).

These analyses indicate that the orbiting body is a planet with high
significance.

\section{Planetary parameters}
\label{sec:planpar}

\begin{deluxetable}{lr}
\tablecaption{
	\label{tab:orb}
	Orbital fit and planetary parameters for the \hatcurb\ system.
}
\tablehead{
	\colhead{Parameter} &
	\colhead{Value}
}
\startdata
\hspace{-7pt} Ephemeris: &\\
Period (day)\tablenotemark{a}	& \hatcurP	\\
$T_{\mathrm{c}}$ (HJD)\tablenotemark{a}& \hatcurT\\
Transit duration (day)		& \hatcurdur	\\
Ingress duration (day)		& \hatcuringdur	\\
\hline
\hspace{-7pt} Orbital parameters: &\\
$\gamma$ (\kms)                 & \hatcurgamma	\\
$K$ (\ms)			& \hatcurK	\\
$e$ \tablenotemark{a}		& $0$	        \\
\hline
\hspace{-7pt} \Lc\ parameters:  &\\
$a/\rstar$		        & \hatcurar	\\
$\rpl/\rstar$			& \hatcurrprstar\\
$b$                             & \hatcurimp	\\
\hline						
\hspace{-7pt} Planetary parameters: &\\
$i$ (deg)                       & \hatcuri      \\
$a$ (AU)            		& \hatcurarel	\\
\mpl\ (\mjup)			& \hatcurm	\\
\rpl\ (\rjup)			& \hatcurr	\\
\rhopl\ (\gcmc)			& \hatcurrho	\\
\gpl\ (\mss)\tablenotemark{b}   & \hatcurg	\\
\teq\ (K)\tablenotemark{c}      & \hatcurTeq    \\
$\Theta$  \tablenotemark{d}     & \hatcurSaf    \\
\fpl\ (10$^9$ erg cm$^{-2}$ s$^{-1}$)\tablenotemark{e}& \hatcurfp
\enddata
\tablenotetext{a}{Fixed in the orbital fit.}
\tablenotetext{b}{Based on only directly observable quantities,
	see \citet[][Eq. 4]{southworth07}.}
\tablenotetext{c}{Planetary thermal-equilibrium surface temperature.}
\tablenotetext{d}{Safronov number, see \citet[][Eq. 2]{hansen07}}
\tablenotetext{e}{Stellar flux at the planet.}
\end{deluxetable}

To determine the \lc\ parameters of \hatcurb\ we fitted the four available
transit \lcs\ simultaneously, as described in \citet{shporer08}.
A circular orbit was assumed.
We adopted a quadratic limb-darkening law for the star,
and determined the appropriate coefficients, $u_1$ and $u_2$, by
interpolating on the
\citet{claret00, claret04} grids for the atmospheric model described above.
For the \woc\ \lc, observed with no filter, we adopted the
Cousins $R$-band limb-darkening coefficients as the CCD response
resembles a ``wide-{\it R}'' filter \citep{brosch08}. The values of the
limb-darkening coefficients used for each \lc\ are listed on \tabr{lcsum}.

The drop in flux in the \lcs\ was modeled with the formalism of
\citet{mandel02}. The five adjusted parameters in the fit were
i) the period $P$;
ii) the mid-transit time of the $N_{tr} = 0$ transit event, $T_{\mathrm{c,0}}$,
denoted simply as $T_\mathrm{c}$;
iii) the relative planetary radius, $r\equiv R_p/\rstar$;
iv) the orbital semi-major axis scaled by the stellar radius, $a/\rstar$;
v) the impact parameter, $b\equiv a\cos(i)/\rstar$, where
$i$ is the orbital inclination angle.

Accounting for correlated noise in the photometric data \citep{pont06}
was done similarly to the ``time-averaging'' method of
\citet[][see also \citealt{shporer08}]{winn08}.
After a preliminary analysis we binned the residual light curves
using bin sizes close to the duration of ingress and egress,
which is 27 minutes.	
The presence of correlated noise in the data was quantified as the ratio
between the standard deviation of the binned residual \lc\ and
the expected standard deviation assuming pure white noise.
This ratio is calculated separately for each bin size and
we defined $\beta$ to be the largest ratio among the bin sizes we used.
For each \lc\ we multiplied
the relative flux errors by $\beta$ and repeated the analysis.
The $\beta$ value of each \lc\ is listed on
\tabr{lcsum}.

We used the Markov Chain Monte Carlo algorithm \citep[MCMC, see,
e.g.][]{holman07} to derive the best fit parameters and their
uncertainties. The MCMC algorithm results in a distribution of each
of the fitted parameters. We took the distribution median to be the
best fit value and the values at 84.13 and 15.87 percentiles to be the
+1\,$\sigma$ and -1\,$\sigma$ confidence limits, respectively.

Our four follow-up \lcs\ are presented in \figr{lc2} where
the fitted model is overplotted and the residuals are also shown.
The result for the radius ratio is $\rpl/\rstar=\hatcurrprstar$, and
the normalized separation is $a/\rstar = \hatcurar$.
From the mass of the star (\tabr{stelpar}), the
orbital parameters (\secr{rv}) and the \lc\ parameters, the physical
planetary parameters (such as mass, radius) are calculated in a
straightforward way, and are summarized in \tabr{orb}.
We note that $a/\rstar$, as derived from the \lc\ fit, is an important
constraint for the stellar parameter determination (\secr{stelpar}),
which in turn defines the limb-darkening coefficients that are used in
the \lc\ analytic fit. Thus, after the initial analytic fit to the
\lcs\ and the stellar parameter determination, we performed another
iteration of the \lc\ fit. We found that the change in parameters was
imperceptible.

\section{Discussion}
\label{sec:disc}

\ifthenelse{\boolean{emulateapj}}{\begin{figure}[t]}{\begin{figure}[t]}
\ifpdf
\plotone{f5.pdf}
\else
\plotone{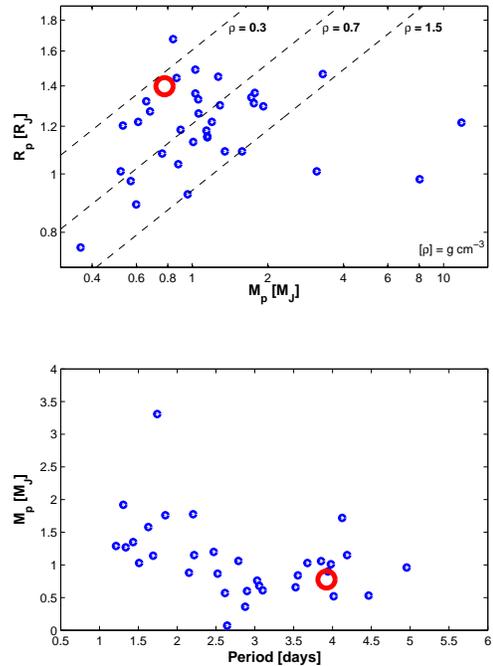}
\fi
\caption{
	{\it Top panel:} The radius-mass diagram for the known transiting
	planets, in log-log scale, with \hatcurb\ marked by an open circle.
	Constant density
	lines are over-plotted. The diagram visually shows that the new
	planet has a relatively low density. The Neptune-mass planet
	GJ~436b, orbiting an M star, is positioned beyond the
	lower-left corner of this diagram.\\
	{\it Bottom panel:} The mass-period diagram for the known transiting
	planets, in linear scale, with \hatcurb\ marked by an open circle.
	Three planets are positioned outside the boundaries of this diagram: 
	The long-period planet \hd{17156b},
	and the massive planets HAT-P-2b (a.k.a.~\hd{147506b}) and XO-3b.
	The planet with the lowest mass in this diagram is GJ~436b, orbiting
	an M star, and the one with the largest mass is CoRoT-Exo-2b.\\
	This figure is based on data taken from
	http://www.inscience.ch/transits/ on 2008 June 1st.
\label{fig:rmp}}
\ifthenelse{\boolean{emulateapj}}{\end{figure}}{\end{figure}}

We present here the discovery of a new transiting extra-solar planet,
\hatcurb, with a mass of \hatcurm\,\mjup, radius of \hatcurr\,\rjup\
and orbital period of \hatcurP\,days. The $V=12.3$\,mag host star is among
the faintest planet-host stars identified with small-aperture,
wide-field, ground-based campaigns. In fact, together with WASP-5
\citep{anderson08} they are currently the faintest planet-host stars
discovered with such campaigns, where the transit depth is below 2\%.

To show the main characteristics of the new planet relative to the
currently known transiting planets we plot in \figr{rmp} the position
of the latter on the radius-mass diagram, on the top panel, and the
mass-period diagram, on the bottom panel. In both panels \hatcurb\
position is marked by an open circle.

The radius-mass diagram,
presented here in log-log scale, visually shows that the new planet
has a low density, of \rhopl = \hatcurrho\,\gcmc, similar to that of
\hd{209458b} \citep[e.g.,][]{knutson07, mazeh00},
WASP-1b \citep{cameron07, shporer07, charbonneau07},
HAT-P-1b \citep{bakos07, winn07b} and CoRoT-Exo-1b \citep{barge08}.

Comparing the measured radius to the modeled radius of \citet{burrows07},
for the same stellar luminosity, semi-major axis and (core-less) planet
mass as the \hatcur\ system, shows the measured value is larger than
the modeled one by 1-2\,$\sigma$. In this comparison we assumed solar
opacity for the planetary atmosphere. As shown by \citet{burrows07},
the transit radius effect together with enhanced planetary atmospheric opacities,
relative to solar opacity, can be responsible for inflating the planetary
radius. Since atmospheric opacities are related to the host star metallicity,
and \hatcur\ metallicity seems to be super-solar, it is likely that these
two effects account for the radius difference, as is the case for WASP-1b
\citep[][see their Fig.~7]{burrows07}. Other mechanisms that may explain the
increased radius involve downward transport and dissipation of kinetic
energy \citep{showman02} and layered convection \citep{chabrier07}.

According to its \teq\ and Safronov number, listed on \tabr{orb},
\hatcurb\ is likely a Class~II planet, as defined by \citet{hansen07}.
This classification is in accordance with its low density, since members
of this class usually have smaller masses and larger radii than those
of Class I.

The stellar incident flux at the planet is
\fpl\ = \hatcurfp\,10$^9$\,erg cm$^{-2}$ s$^{-1}$,
which is at the lower end of the \fpl\ range for pM planets, according
to the pL/pM planet classification of \citet{fortney08b}.
Since it is positioned near the transition region in \fpl\ between these
two classes of planets it may be used to further characterize them.

In the mass-period diagram the new planet is positioned near XO-1b
\citep{mccullough06, holman06}, OGLE-TR-182b \citep{pont08}, OGLE-TR-111b
\citep{pont04, winn07a} and HAT-P-6b \citep{noyes08}.
The position of \hatcurb, along with that of neighbouring planets,
suggests that the mass-period relation \citep{mazeh05, gaudi05},
i.e., the decrease of planetary mass with increasing orbital period,
levels-off at periods $\gtrsim 3$\,days.

The ephemeris obtained here is based on the four follow-up \lcs.
We compared it to the mid-transit time of the HATNet \lc, obtained some
1000\,days earlier, by propagating the error on the period. The result
shows a difference of about 1.5\,$\sigma$ between the mid-transit times of
the follow-up \lcs\ and that of the HATNet \lc. Future follow-up \lcs\
will be able to study more thoroughly this possible discrepancy.

Finally, we note that due to the line-of-sight stellar rotation velocity
of \hatcurvsini\,\kms\ and the non-zero impact parameter
\hatcurimp, this planet is a good candidate for the
measurement of the Rossiter-Mclaughlin effect \citep[e.g.,][]{winn05}.
Eq.~6 of \citet{gaudi07} gives an expected amplitude for the effect of
more than 100\,\ms, which is larger than the orbital amplitude.
However, since this is a relatively faint star, with $V=12.3$\,mag,
such a measurement will be challenging.


\acknowledgments

Operation of the HATNet project is funded by NASA grants
NNG04GN74G and NNX08AF23G.
These observations have been partially funded by the Optical
Infrared Coordination network (OPTICON),
a major international collaboration supported by the
Research Infrastructures Programme of the
European Commission’s Sixth Framework Programme.
TM acknowledges support from the ministry of science culture \& sport
through a grant to encourage French-Israeli scientific collaboration.
GB is a National Science Foundation Fellow, under grant AST-0702843.
We acknowledge partial support from the Kepler Mission under NASA
Cooperative Agreement NCC2-1390 (DWL, PI).
GK thanks the Hungarian Scientific Research
Fund (OTKA) support through grant K-60750.
GT acknowledges partial
support for this work from NASA Origins grant NNG06GH69G.



\end{document}